\documentstyle[preprint,aps]{revtex}

\begin{document}
\draft

\title{Medium Effects on Binary Collisions with the $\bf{\Delta}$ Resonance}

\author{T.-S. H. Lee}
\address{\underline{Physics Division, Argonne National Laboratory, 
Argonne, IL. 60439-4843}}

\maketitle

\begin{abstract}
To facilitate the relativistic heavy-ion calculations based on transport 
equations, the binary collisions involving a $\Delta$ resonance
in either the entrance channel or the exit channel
are investigated within a Hamiltonian formulation of $\pi NN$ interactions.
An averaging procedure is developed to define a quasi-particle $\Delta^*$ and
to express the experimentally measured
$NN\rightarrow \pi NN$ cross section in terms of an effective $NN\rightarrow
N\Delta^\ast$ cross section. In contrast to previous works, the 
main feature of the present approach is that the
mass and the momentum of the
produced $\Delta^*$'s are calculated dynamically from 
the bare $\Delta \leftrightarrow \pi N$
vertex interaction of the model Hamiltonian and are constrained by the
unitarity condition.
The procedure is then extended to define the
effective cross sections for the experimentally inaccessible $N\Delta^\ast \rightarrow
NN$ and $N\Delta^\ast \rightarrow N\Delta^\ast$ reactions.  The predicted cross
sections are significantly different from what are commonly assumed in
relativistic heavy-ion calculations.  The $\Delta$ potential in nuclear matter has
been calculated by using a Bruckner-Hartree-Fock approximation.  
By including the mean-field effects
on the $\Delta$ propagation, the effective cross sections
of the $NN\rightarrow N\Delta^\ast$, $N\Delta^\ast \rightarrow NN$ and 
$N\Delta^\ast \rightarrow N\Delta^\ast$ reactions in nuclear matter are
predicted. It is demonstrated that the density dependence is most dramatic in the
energy region close to the pion production threshold.
\end{abstract}

\bigskip
\pacs{Pacs Numbers: 25.75.+r}

\pagebreak

\section{Introduction}

Nuclear matter with a high population of nucleon resonances ($\Delta, N^\ast$'s) 
is expected to
exist in some astrophysical objects and can be created in relativistic heavy-ion
collisions.  With the very dedicated experimental efforts at GSI, AGS and CERN 
in the past decade,
it appears that the properties of such a nuclear system can now be investigated.
Experimental evidence for creating a $\Delta$-rich nuclear
system during relativistic heavy-ion collisions 
have recently been reported \cite{mosela,gsi}. 
This interesting development has raised two theoretical issues.
First, it is necessary to examine the extent to which the theoretical models
employed in identifying the $\Delta$-rich matter are valid. In the calculations 
based on transport equations \cite{bertsch1,bertsch2,mosel2,mosel3,li,ko,muntz}, some 
assumptions were made to define the cross sections of binary collisions involving 
a $\Delta$ in either the entrance channel 
or the exit channel. This needs to be clarified within a rigorous formulation of the 
scattering theory with resonance excitations. The medium effects on $\Delta$
propagation assumed in those calculations must also be examined from a
more fundamental point of view.  One possibility is to calculate the $\Delta$ mean
field using the well-developed nuclear many-body methods \cite{mahaux,lee4}.
The second theoretical issue is how to develop a microscopic approach 
to understand the $\Delta$-rich matter in terms of the elementary
$NN$ , $N\Delta$ and $\Delta\Delta$ collisions and their coupling with
the pion production channels. These two theoretical issues are closely related
and must be addressed within the same theoretical framework. The heart of the 
problem is to define precisely and to determine quantitatively all possible 
binary collisions in the $\Delta$-rich matter. In this paper, we will address this 
problem by employing a Hamiltonian model of $\pi NN$
interactions which was previously developed to 
describe $NN$ scattering up to 2 GeV \cite{lee1,lee2,lee3} and the 
$\Delta$-nucleus dynamics in pion-nucleus reactions \cite{lee4}. 
Our main objective is to 
provide the theoretical input to the relativistic heavy-ion calculations using 
transport equations \cite{bertsch1,bertsch2,mosel2,mosel3,li,ko,muntz}.
 
In section II, we briefly review the coupled $NN\bigoplus N\Delta$ scattering equations
within the $\pi NN$ Hamiltonian model 
developed in Refs.\ \cite{lee1,lee2,lee3}. 
We derive formulas in section III to relate the experimentally
observed $NN\rightarrow \pi NN$ reaction to the resonating $N\Delta$ state.
An averaging procedure is then developed to define a quasi-particle $\Delta^\ast$
and to evaluate the cross sections
for binary collisions with a $\Delta^\ast$ in either the entrance channel or the
exit channel. In section IV, we recall the approach of Ref.\ \cite{lee4}
to calculate the mean-field effects on
$\Delta$ propagation which are then used to calculate
the medium-corrected cross sections.
In section V, we present our predictions of the 
cross sections for $NN\rightarrow N\Delta^\ast$, $N\Delta^{\ast} \rightarrow NN$ and 
$N\Delta^\ast \rightarrow N\Delta^\ast$ transitions in free space and in nuclear medium.
A summary is given in section VI.

\section{Hamiltonian Model of $\bf \pi NN$ interactions}

An important advance in intermediate energy physics is the development of
a microscopic approach to understand the
nuclear reactions induced by pions, protons, and electrons in terms of the
interactions between $\pi$, $N$, and $\Delta$ degrees of freedom. 
Such an approach must start with a $\pi NN$ model which can describe
the following elementary processes
\begin{eqnarray}
\pi N &\rightarrow &\pi N,\,\,\,\,\,\,\,\,\,\,\,\,\,E_L \leq \,\,300\,\, \rm{MeV} \,; \\ 
NN &\rightarrow & NN\,,
\nonumber \\
&\rightarrow  &\pi NN,\,\,\,\,\,\,\,\,\,E_L \leq 1000\,\, \rm{MeV} \,; \\
\pi d &\rightarrow & \pi d\,,
\nonumber \\
&\rightarrow & NN\,,
\nonumber \\
&\rightarrow &\pi NN\,\,\,\,\,\,\,\,\,\,\,\,\,\,E_L \leq \,\,300 \,\, \rm{MeV} \,.
\end{eqnarray}
The developments of $\pi NN$ models have been reviewed in 
Ref.\ \cite{garcilazo}. In this work we employ the
$\pi NN$ model developed in Ref.\ \cite {lee1,lee2,lee3}.
The main feature of the 
constructed $\pi NN$
Hamiltonian is that the $NN \leftrightarrow N\Delta$ interaction can be directly
constrained by data on the $NN \rightarrow N\Delta \rightarrow \pi NN$ reaction, and
the $N\Delta \rightarrow N\Delta$ interaction can also be related to the $\pi d$ reactions.  
This allows a more realistic determination of the phenomenological parts of the interactions
involving the $\Delta$.  For example, it was found \cite{lee1,lee2,lee3} that
the ranges of the $\pi NN$ and $\pi N\Delta$
vertices in the $NN \leftrightarrow N\Delta$ transition potential
must be less than about 750 MeV for a monopole form if the
the $NN \rightarrow \pi NN$ reaction cross sections can be reasonably described. 
This phenomenological procedure is not avoidable in any meson-exchange model 
since a fundamental theory of short-range $NN$ and 
$N\Delta$ interactions is still not available. 

In this first application of the $\pi NN$ model to relativistic heavy-ion
collisions, we will focus on the $\Delta$ resonance and neglect the weaker non-resonant
$\pi N$ interactions. The $\pi NN$ model Hamiltonian within the 
formulation of Refs.\ \cite{lee1,lee2,lee3} can then be written as

\begin{equation}
H = H_0 + H_{int}\, ,
\end{equation}
where $H_0$ is the sum of free energy operators for 
the $N,\,\Delta$ and $\pi$ degrees of
freedom, and 

\begin{eqnarray}
H_{int} &=& \sum^{2}_{i=1} 
\left[ h_{\pi N,\Delta} (i) + h_{\Delta ,\pi N} (i) \right]
\nonumber \\
&+& {1 \over 2} \sum^{2}_{i,j=1} \left[ V_{NN,NN} (i,j) +
        V_{NN,N\Delta} (i,j) + V_{N\Delta,NN} (i,j) +V_{N\Delta,N\Delta}(i,j) \right]
\end{eqnarray}
Note that the vertex interactions $h_{\pi N,\Delta}$ and $h_{\Delta ,\pi N}$
describe the $\pi N \leftrightarrow \Delta$
transition.  They can renormalize the bare $\Delta$ mass 
and generate a $N\Delta \rightarrow \Delta N$ interaction due to the exchange of
an ``on-mass-shell" pion. In the following presentation, this
$N\Delta \rightarrow \Delta N$ is included in $V_{N\Delta,N\Delta}$ of Eq.\ (5).

The derivation of the $\pi NN$ scattering 
equations from the above model Hamiltonian can be found in
section II of Ref.\ \cite{lee2}.
By some straightforward derivations using
Eqs.\ (3.27), (3.31), and
(3.24) of Ref.\ \cite{lee2}, we obtain the following equations defined in
the coupled $NN \oplus N\Delta$ space
\begin{eqnarray}
T_{NN,NN} (E) &=& \hat{V} _{NN,NN} (E) + \hat{V} _{NN,NN} (E)
G_{NN}(E)T_{NN,NN} (E) \,, \\
T_{N\Delta ,NN} (E) &=& \Omega^{(-)+}_{N\Delta}(E)V_{N\Delta ,NN} \left[ 1 + 
G_{NN}(E)T_{NN,NN} (E) \right]\, , \\
T_{NN,N\Delta} (E) &=& \left[ 1 + T_{NN,NN} (E) G_{NN}(E) \right]
V_{NN,N\Delta}\Omega^{(+)}_{N\Delta}(E) \, , \\
T_{N\Delta,N\Delta}(E)&=& t_{N\Delta,N\Delta}(E) + T_{N\Delta,NN}V_{NN,N\Delta}
\Omega^{(+)}_{N\Delta}(E) \, .
\end{eqnarray}
In the above equations, we have defined the propagators
\begin{eqnarray}
G_{NN}(E)&=&\frac{P_{NN}}{E-H_0+i\epsilon} \,, \\
G_{N\Delta}(E)&=&\frac{P_{N\Delta}}{E-H_0-\Sigma(E)} \, ,
\end{eqnarray}
where the $\Delta$ self-energy is determined 
by the $h_{\pi N \leftrightarrow \Delta}$ vertex interaction in the presence of a
spectator nucleon
\begin{equation}
\Sigma _{\Delta} (E) = \sum^{2}_{i=1} h_{\Delta ,\pi N} (i)\,
{P_{\pi NN} \over E - H_0 + i\epsilon}\,
h_{\pi N,\Delta} (i)\, .
\end{equation}
$P_{NN}$, $P_{N\Delta}$, and 
$P_{\pi NN}$ are respectively the projection operators for the $NN$, $N\Delta$, and 
$\pi NN$ states.  

The main feature of the above scattering formulation is that the effects due to
the $V_{N\Delta,N\Delta}$ interaction are isolated in the
$N\Delta$ scattering t-matrix $t_{N\Delta,N\Delta}$ defined by
\begin{eqnarray}
t_{N\Delta,N\Delta}(E) = V_{N\Delta,N\Delta} + V_{N\Delta,N\Delta}G_{N\Delta}(E)
t_{N\Delta,N\Delta}(E) \, ,
\end{eqnarray}
and the $N\Delta$ scattering operators defined by
\begin{eqnarray}
\Omega^{(+)}_{N\Delta}(E) = 1 + G_{N\Delta}(E) t_{N\Delta,N\Delta}(E) \, , \\
\Omega^{(-)+}_{N\Delta}(E) = 1+ t_{N\Delta,N\Delta}(E)G_{N\Delta}(E) \, .
\end{eqnarray}
The $N\Delta$ scattering also has a contribution to the effective $NN$ potential
in Eq.\ (6). Explicitly, we have 
\begin{equation}
\hat{V}_{NN,NN} (E) = V_{NN,NN} + U^{(2)}_{NN,NN} (E)\, ,
\end{equation}
with
\begin{equation}
U^{(2)}_{NN,NN} (E) = V_{NN,N\Delta}G_{N\Delta}(E)\left[ 1 +
t_{N\Delta,N\Delta}(E)G_{N\Delta}(E) \right] V_{N\Delta ,NN} \, .
\end{equation}

The task is then to find an appropriate parameterization of
the model Hamiltonian Eq.\ (5) to best reproduce the data of the $\pi NN$ 
processes listed in Eqs.\ (1)-(3). This was first achieved by Betz and Lee \cite{betz} 
using the separable parameterization.
In this work we consider the meson-exchange parameterization of
Refs.\ \cite{lee1,lee2,lee3}. It was found that the $\pi NN$ data can be best
reproduced by using: (1) the vertex 
interaction $h_{\pi N\leftrightarrow \Delta}$ determined
from fitting the $\pi N$ scattering phase shift in $P_{33}$ channel, (2) 
the one-pion-exchange model \cite{gari} of the
transition potentials $V_{NN \leftrightarrow N\Delta}$ and $V_{N\Delta,N\Delta}$
with a monopole form factor of a cutoff parameter $\Lambda$= 650 MeV/c (3),
the $NN$ interaction $V_{NN,NN}$ of Eq.\ (16) is defined by using a subtraction of 
the Paris potential \cite{paris}
\begin{equation}
V_{NN,NN} = V_{Paris} - U^{(2)}_{NN,NN} (E = E_s) \, ,
\end{equation}
with $E_s=10$ MeV laboratory energy.
Note that the above definition of the $NN$
interaction amounts to removing phenomenologically the two-pion-exchange with an
intermediate $N\Delta$ state from the Paris potential, in order to avoid the double
counting of the $N\Delta$ effect.  This approach was also developed
independently by Sauer and collaborators \cite{sauer}.
The details of the determination of the model 
Hamiltonian as well as comparisons with the $\pi NN$ data 
can be found in Refs.\ \cite{lee1,lee2,lee3}. Here we illustrate the
validity of the model by showing in Fig.\ 1 that 
the predictions of the model are also in good agreement
with the most recent np polarization data \cite{spinka}.

Before we proceed further, it is important to clarify the meaning of the $\Delta$
degree of freedom of the considered $\pi NN$ model. Because of the presence
of the vertex interaction $h_{\Delta\leftrightarrow\pi N}$, the $\Delta$ 
is certainly not a physical particle which can be detected experimentally.
To investigate the $\Delta$ dynamics, we need to consider
the $NN \leftrightarrow \pi NN$ and $\pi NN \rightarrow
\pi NN$ reactions. The amplitudes of these reactions can be calculated 
from the vertex
interaction $h_{\Delta\leftrightarrow\pi N}$ and the baryon-baryon t-matrix 
defined in Eqs.\ (6)-(9)  by the following equations
\begin{eqnarray}
T_{NN, \pi NN}(E) &=& T_{NN,N\Delta}(E)G_{N\Delta}
\left[ \sum^{2}_{i=1}h_{\Delta,\pi N}(i) \right] \, , \\
T_{\pi NN,NN}(E)&=&\left[\sum^{2}_{i=1}h_{\pi N,\Delta}(i)\right]
G_{N\Delta}(E)T_{N\Delta,NN}(E) \, ,\\
T_{\pi NN, \pi NN}(E) &=&\left[\sum^{2}_{i=1}h_{\pi N,\Delta}(i)\right]
G_{N\Delta}(E)T_{N\Delta,N\Delta}(E)\left[ \sum^{2}_{i=1}h_{\Delta,\pi N}(i)\right] \, .
\end{eqnarray}

\section{Effective binary cross sections with $\bf\Delta^*$}

We will only consider the transport equations of relativistic heavy-ion 
collisions which are determined by the cross sections of the
 collisions between ``on-mass-shell" physical particles.
Within the model defined by the $\pi NN$ Hamiltonian Eqs.\ (4)-(5),
the transport equations can then only be written in terms
of $\pi$ and $N$ variables. It is however possible to express the dynamics
associated with the pion in terms of 
a quasi-particle $\Delta^\ast$ which has a mass equal to 
the invariant mass of a $\pi N$ subsystem.
The transport equations can then be re-written in terms of $N$ and $\Delta^\ast$.
In such a approach, the mass of the $\Delta^\ast$ can range from $m_N+m_\pi$
to the highest energy allowed by the collision energy. 
To carry out calculations, it is necessary to define the 
$NN \leftrightarrow N \Delta^\ast$ and $N \Delta^\ast \rightarrow N\Delta^\ast$ 
cross sections. In the following we will try to define rigorously these cross sections
in terms of the $NN\leftrightarrow \pi NN$ and $\pi NN \leftrightarrow \pi NN$ 
amplitudes defined in the previous section.

To proceed, let us first consider the $NN \rightarrow \pi NN$ and 
the three-body kinematics in the center of mass frame (Fig.\ 2):
\begin{eqnarray}
\vec{k} &=& \vec{k}_1 \, , \nonumber \\
        &=&  -(\vec{k}_2+\vec{k}_3) \, .
\end{eqnarray}
For a given total collision energy $E$ and a given momentum $\vec{k}$, 
the invariant mass
$M$ and the internal relative momentum $\vec{q}$ of the $\pi N$(23) 
subsystem, called $\Delta^\ast$, 
are defined by the following relations
\begin{eqnarray}
E&=&E_N(k) + [ M^2+k^2]^{1/2} \, , \\
M&=&E_N(q)+E_\pi(q) \, , 
\end{eqnarray}
where $E_\alpha(q) =(m_\alpha^2+q^2)^{1/2}$ is the energy of a particle with
mass $m_\alpha$. Eq.\ (23) also leads to a useful relation
\begin{eqnarray}
M^2 = E^2 - 2EE_N(k) + m_N^2 \, .
\end{eqnarray}
Clearly, Eqs.\ (24) and (25) imply that the invariant mass of the $\Delta^\ast$ 
is restricted to the 
region $(m_N+m_\pi) \le M \le (E-m_N)$.
By using Eqs.\ (23)-(24), the momenta $k$ and $q$ can be written as
\begin{eqnarray}
k &=&\frac{1}{2E}[(E^2-m_{N}^2-M^2)^2-4M^2m_{N}^2]^{1/2} \, ,\\
q &=&\frac{1}{2M}[(M^2-m_N^2-m_\pi^2)^2-4m_{N}^2m_{\pi}^2]^{1/2} \, . 
\end{eqnarray}

In terms of the kinematical variables defined above and in Fig.\ 2, the matrix
element of the $NN \rightarrow \pi NN$
transition operator, Eq.\ (20), can be written (suppressing spin-isospin indices) as
\begin{eqnarray}
< \vec{k},\vec{q} \mid T_{\pi NN, NN}(E) | \vec{k}_0 >
= h_{\pi N,\Delta}(\vec{q})\frac{1}{E-E_N(k)-E_{\Delta}(k) -\Sigma(E,\vec{k})}
T_{N\Delta,NN}(\vec{k},\vec{k}_0,E) \, .
\end{eqnarray}

We choose the normalization $<\vec{k}^\prime \mid \vec{k} > =\delta(\vec{k}^\prime
-\vec{k})$. The scattering t-matrix is related to the S-matrix by (in operator form)
\begin{eqnarray}
S(E) = 1 - 2\pi i \delta(E-H_0) T(E) \, .
\end{eqnarray}
We next define the following partial-wave expansions 
\begin{eqnarray}
h_{\pi N, \Delta }(\vec{q}) = y^{s_\Delta m_{s_\Delta}}_{l_{23}s_{23}}(\hat{q})
h_{\pi N, \Delta}(q) < s_{\Delta} m_{s_{\Delta}} \mid \, ,
\end{eqnarray}
and 
\begin{eqnarray}
T_{N\Delta,NN}(\vec{k},\vec{k}_0,E)= 
\sum_{J,M_J}\sum_{LS,L_\Delta S_\Delta} y^{J M_J}_{L_\Delta S_\Delta}(\hat{k})
 T^{J}_{L_\Delta S_\Delta,LS}(k,k_0,E) y^{JM_J^+}_{LS}(\hat{k}_0) \,.
\end{eqnarray}
In the above equations, we have defined the spin-angular vector 
\begin{eqnarray}
y^{jm}_{ls}(\hat{p}) &=&\sum_{m_lm_s}\sum_{m_{s_1}m_{s_2}}Y_{lm_l}(\hat{p})
\mid s_1 m_{s_1} > \mid s_2 m_{s_2} >
\nonumber \\
&\times& <lsm_lm_s|jm><s_1s_2m_{s_1}m_{s_2}|sm_s> \, ,
\end{eqnarray}
where $Y_{lm_l}(\hat{q})$ is the usual spherical harmonic function, 
$(s_i , m_{s_i}) $ are the spin variables for the i-th particle.

By using the definitions Eqs.\ (28)-(29) and 
the variables defined by Eqs.\ (23)-(27), we can write the
total cross section of the two-step process 
$NN \rightarrow N\Delta \rightarrow NN\pi$ (Fig.\ 2) as
\begin{eqnarray}
\sigma^{tot}(E)&=& \frac{1}{(2s'_1 +1)(2s'_2 +1)}
\sum_{spins}\frac{(2\pi)^4}{k_0^2}
\left[ \frac{k_0E_{N}(k_0)}{2} \right]
\int d\vec{k} \int d\vec{q} \nonumber \\ 
& &\delta (E - E_N(k) - \sqrt{(E_N(q)+E_\pi(q))^2+k^2}) 
\mid \sum_{m_{s_{\Delta}}}
< m_{s_2}m_{s_3}  \mid h_{\pi N,\Delta}(\vec{q})
\mid m_{s_\Delta} > \nonumber \\
&\times&\frac{1}{E-E_N(k)-E_{\Delta}(k)-\Sigma(k,E)}
<m_{s_{\Delta}} m_{s_1} \mid T_{N\Delta,NN}(\vec{k},\vec{k}_0,E) \mid 
m_{s_1}^\prime m_{s_2}^\prime > \mid ^2 \, ,
\end{eqnarray}
where the initial momentum $k_0$ is defined by the total energy 
$E=E_{N}(k_0)+E_{N}(k_0)$. The $\Delta$ self-energy is found to be
\begin{eqnarray}
\Sigma(k,E) = \int\frac{|h_{\Delta, \pi N}(q')|^2 q^{\prime 2} d q'}
{[(E-E_N(k))^2-k^2]^{1/2}-E_N(q')-E_\pi (q')+i\delta} \, .
\end{eqnarray}

Substituting the expansions Eqs.\ (30)-(31) into Eq.\ (33) and 
integrating over angles $\hat{q}$ and $\hat{k}$, we obtain
\begin{eqnarray}
\sigma^{tot}_{NN\pi\leftarrow NN}(E) 
= \sum_{J,LS,L_{\Delta}S_{\Delta}} d\sigma^{J}_{L_\Delta
S_\Delta,LS}(E),
\end{eqnarray}
with 
\begin{eqnarray}
d\sigma^{J}_{L_{\Delta} S_{\Delta} ,LS}(E)=\frac{4\pi}{k_0^2} (J+\frac{1}{2})
\int_{0}^{k_{max}}dk\mid N_{N\Delta}^{1/2}(k)
T^{J}_{L_{\Delta} S_{\Delta} ,LS}(k,k_0,E)\rho_{NN}^{1/2}(E ) \mid ^2 \, ,
\end{eqnarray}
where $k_{\max}$ can be calculated from Eq.\ (26) by setting $M = m_{\pi} + m_N$, and 
\begin{eqnarray}
\rho_{NN}(E )=\pi \frac{k_0 E_{N}(k_0 )}{2} \, ,
\end{eqnarray}
\begin{eqnarray}
N_{N\Delta}(k)=\pi k^2\frac{qE_N(q)E_\pi(q)E_M(k)}{M^2} \mid 
\frac{h_{\Delta \leftrightarrow \pi N}(q)}
{E-E_N(k)-E_{\Delta}(k)-\Sigma(k,E)} \mid ^2 \, .
\end{eqnarray}

To have a better understanding of Eq.\ (36), we use Eq.\ (25) to
change the integration variable
\begin{eqnarray}
dk =-\frac{ME_{N}(k)}{kE}dM \, .
\end{eqnarray}
Eq.\ (36) can then be written as an integral over the invariant mass $M$ of the
$\pi N$ subsystem
\begin{eqnarray}
d\sigma^{J}_{L_\Delta S_\Delta ,LS}=\frac{4\pi}{k_0^2} (J+\frac{1}{2})
\int_{m_{\pi}+m_N}^{E-m_N}dM\mid \Gamma_{N\Delta}(M)^{1/2}
T^{J}_{L_\Delta S_\Delta ,LS}(k,k_0,E)\rho_{NN}^{1/2}(E) \mid ^2 \, .
\end{eqnarray}
where 
\begin{eqnarray}
\Gamma_{N\Delta}(M)&=&\pi \left[ \frac{k E_N(k)E_M(k)}{E} \right]
\left[ \frac{qE_N (q)E_{\pi}(q)}{M} \right]
\nonumber \\
&\times&\mid \frac{h_{\Delta , \pi N}(q)}
{E-E_{N}(k)-E_{\Delta}(k)-\Sigma(k,E)} \mid ^2 \, .
\end{eqnarray}
Note that in evaluating the above integral, we need to use Eqs.\ (26) and (27) to
express $k$ and $q$ in terms of the invariant mass $M$ and the total
energy $E$. By Eqs.\ (26) and (34), the self-energy $\Sigma(k,E)$ is also   
a function of the total energy $E$ and the invariant mass $M$.

Equation (40) is an exact expression. The integrand contains the 
dependence on the invariant mass $M$ of the $\Delta^\ast$, 
$d\sigma^J_{L_\Delta S_\Delta,LS}/d M$,  which is needed for
a detailed transport-equation calculation.
For practical applications and making contact with the current calculations of
relativistic heavy-ion collisions, it is interesting 
to develop an averaging
procedure such that the cross section for a given total
energy E can be written in terms of only one averaged $\Delta^*$. This is our next task.

We use Eqs.\ (26) and (41) to define an average relative momentum $\bar{k}$ 
of the $N\Delta^\ast$
two-body system
\begin{eqnarray}
\bar{k} = \frac{\int_{m_N + m_{\pi}}^{E-m_N} \Gamma_{N\Delta}(M)kdM}
{\int_{m_N + m_{\pi}}^{E-m_N}\Gamma_{N\Delta}(M)dM} \, ,
\end{eqnarray}
The corresponding average mass of $\Delta^\ast$ is then calculated from Eq.\ (25)
\begin{eqnarray}
\bar{M}^2=E^2-2EE_N(\bar{k})+m_N^2 \, .
\end{eqnarray} 
Assuming that the transition t-matrix in the 
integrand of Eq.\ (40) can be calculated at the 
average momentum $\bar{k}$,
we then obtain the following ``factorized" form 
\begin{eqnarray}
\sigma^{tot}_{N\Delta^{\ast}\leftarrow NN}= \sum_{J,L_\Delta S_\Delta,LS}
d\bar{\sigma}^J_{L_\Delta S_\Delta,LS} \, ,
\end{eqnarray}
with
\begin{eqnarray}
d\bar{\sigma}^{J}_{L_\Delta S_\Delta ,LS}=\frac{4\pi}{k_0^2} (J+\frac{1}{2})
\mid \rho_{N\Delta}^{1/2}(E)
T^{J}_{L_\Delta S_\Delta ,LS}(\bar{k},k_0,E)\rho_{NN}^{1/2}(E) \mid ^2 \, .
\end{eqnarray}
Here we have introduced a phase-space factor for the
$N\Delta^\ast$ two-body state
\begin{eqnarray}
\rho_{N\Delta}(E) = \int_{m_N+m_\pi}^{E-m_N} dM \Gamma_{N\Delta}(M) \, .
\end{eqnarray}
Equation (45) has the form similar to that introduced in the 
literature \cite{bertsch2,mosel2}.  
Here we have explicitly shown how it can be derived from the Hamiltonian
formulation \cite{lee1,lee2,lee3} of 
the $\Delta$ excitation by using the ``factorization approximation".  The
accuracy of this approximation will be examined in section V.\,\, 
We emphasize here that that $\Gamma_{N\Delta}(M)$, defined in Eq.\ (41),
is the consequence of a unitary formulation of the $\pi NN$ processes listed
in Eqs.\ (1)-(3). Equation (46) is significantly different from the parameterizations
of the $N\Delta$ phase-space introduced in Refs.\ \cite{bertsch2,mosel2}. 

We note that Eq.\ (45) has
the usual form for the binary collisions involving only stable particles.
For example, the exact expression for $NN \rightarrow NN$ process can be
obtained by replacing $\rho_{N\Delta}(E)$
in Eq.\ (45) by $\rho_{NN}(E)$.

By using the $N\Delta^\ast$ representation of the $\pi NN$ state, 
it is straightforward to carry out similar derivations 
to express the $\pi NN \rightarrow NN$ and $\pi NN \rightarrow \pi NN$
cross sections in terms of $N\Delta^\ast \rightarrow NN$ and $N\Delta^\ast \rightarrow
N\Delta^\ast$ cross sections. If the averaging procedure defined in Eqs.\ (42) and (46)
is also applied, we obtain the following expressions
\begin{eqnarray}
\sigma^{tot}_{NN\leftarrow N\Delta^\ast}= \sum _{J,LS,L_\Delta S_\Delta}
d\bar{\sigma}^J_{LS,L_\Delta S_\Delta} \, 
\end{eqnarray}
with
\begin{eqnarray}
d\bar{\sigma}^J_{LS,L_\Delta S_\Delta}&=&\frac{4\pi}{\bar{k}^2}
\mid\rho^{1/2}_{NN}(E)
T^J_{LS,L_\Delta S_\Delta}(k_0,\bar{k},E)\rho^{1/2}_{N\Delta}(E)\mid^2 \, , 
\end{eqnarray}
for the $N\Delta^\ast \rightarrow NN$ reaction, and
\begin{eqnarray}
\sigma^{tot}_{N\Delta^\ast\leftarrow N\Delta^\ast}
=\sum_{J,L^\prime_\Delta S^\prime_\Delta,L_\Delta S_\Delta}
d\bar{\sigma}^J_{L_\Delta' S_\Delta', L_\Delta S_\Delta} \, ,
\end{eqnarray}
with
\begin{eqnarray}
d\bar{\sigma}^{J}_{L_\Delta' S_\Delta' ,L_\Delta S_\Delta}
&=&\frac{4\pi}{\bar{k}^2} (J+\frac{1}{2})
\mid \rho^{1/2}_{N\Delta}(E)
T^{J}_{L_\Delta' S_\Delta ' ,L_\Delta S_\Delta}
(\bar{k},\bar{k},E)\rho_{N\Delta}^{1/2}(E) \mid ^2 \, ,
\end{eqnarray}
for the $N\Delta^\ast \rightarrow N\Delta^\ast$ reaction. 

We can further extend the above formula to calculate the differential
cross sections involving a $\Delta^\ast$ resonance in either the entrance channel
or the exit channel. Specifically,
we can define a general scattering amplitude
\begin{eqnarray}
\tilde{T}_{\alpha\beta}(\vec{k}_\alpha,\vec{k}_\beta, E) 
&=& \sum_{J M_J, L_\alpha S_\alpha,
L_\eta S_\beta} y^{JM_J}_{L_\alpha S_\alpha}(\hat{k}_\alpha)
y^{JM_J^+}_{L_\beta S_\beta}(\hat{k}_\beta) \nonumber \\
&\times& \rho^{1/2}_{\alpha}(E)
T^{J}_{L_\alpha S_\alpha, L_\beta S_\beta}(\bar{k}_\alpha,\bar{k}_\beta,E)
\rho^{1/2}_{\beta}(E) \, ,
\end{eqnarray}
where $\alpha,\beta$ can be $NN$ or $N\Delta^\ast$ states, and
$|\vec{k}_\alpha| = \bar{k}, k_0$ for $\alpha= N\Delta^\ast ,NN$.
The differential cross sections are then of the following form
\begin{eqnarray}
\frac{d\sigma_{\alpha\beta}(E)}{d\Omega} = \frac{16\pi^2}{\bar{k}_\beta^2}
\frac{1}{(2s_{1_{\beta}}+1)(2s_{2_{\beta}}+1)}\sum_{spins}
\mid < m_{1_{\alpha}}m_{2_{\alpha}} \mid \tilde{T}_{\alpha\beta}(\vec{k}_\alpha,
\vec{k}_\beta, E) \mid m_{1_{\beta}}m_{2_{\beta}} > \mid^2
\end{eqnarray}
It is easy to show that the total
cross sections Eqs.\ (44), (47), and (49) 
can be obtained by
integrating Eq.\ (52) over the scattering angle $\Omega$.

\section{Medium Effects on the $\bf{\Delta}$ propagation}

We follow the approach developed in Ref.\ \cite{lee4} to calculate
the medium effects on the $\Delta$ propagation. The resulting $\Delta$ potential
is then used to calculate the medium effects on the scattering cross sections
defined in the previous section.

In the one-hole-line approximation, the propagator of a $\Delta$ with momentum
$\vec{p}_{\Delta}$ and energy $\epsilon_{\Delta}$ in nuclear matter
can be written as

\begin{eqnarray}
G_{\Delta}(\vec{p}_{\Delta}, \epsilon_{\Delta})
=\frac{1}{\epsilon_{\Delta} - E_{\Delta}(\vec{p}_{\Delta}) 
- \Sigma^{(1)}(\vec{p}_{\Delta},\epsilon_{\Delta})
-\Sigma^{(2)}(\vec{p}_{\Delta},\epsilon_{\Delta})} \, ,
\end{eqnarray}
with
\begin{eqnarray}
\Sigma^{(1)}(\vec{p}_\Delta,\epsilon_\Delta) = \int
\frac{Q_{\pi N}(k, p_\Delta)\mid h_{\pi N \leftrightarrow \Delta}(k)\mid ^2 k^2 dk}
{\epsilon_\Delta - \sqrt{(E_N (k)+E_\pi (k))^2+p_\Delta ^2} + i\epsilon} \, ,
\end{eqnarray}
where $Q_{\pi N}(k,p_\Delta)$ is an angle-average Pauli operator \cite{lee4} 
for the $\pi N$ state. 
The medium effect on the $\Delta$ propagation is calculated from the $N\Delta$
G-matrix 
\begin{eqnarray}
\Sigma^{(2)}(\epsilon_{\Delta},\vec{p}_{\Delta}) = 
\int_{p \le p_F}  d\vec{p}<\vec{p}_{\Delta} \vec{p}
\mid G_{N\Delta,N\Delta}(W=(\epsilon_{\Delta} + \epsilon_{N}(\vec{p})))
\mid \vec{p}_{\Delta} \vec{p} > \, ,
\end{eqnarray}
where $p_F$ is the Fermi momentum of nuclear matter.

It is convenient to calculate the G matrix in partial-wave representation.
For symmetric nuclear matter( the total isospin and the total angular momentum
are both equal to zero), we find that
\begin{eqnarray}
\Sigma^{(2)}(\vec{p}_{\Delta},\epsilon_{\Delta}) &=& \frac{1}{(2 s_{\Delta}+1)
(2\tau_{\Delta}+1)}\sum_{L_\Delta S_\Delta J T}\frac{(2J+1)(2T+1)}{4\pi} \nonumber \\
&\times&\int_{p \le p_F} d\vec{p} G^{JT}_{L_\Delta S_\Delta ,L_\Delta S_\Delta}
(q,q ,\omega (\epsilon_{\Delta},
\vec{p}_{\Delta},\vec{p}),P ) \, ,
\end{eqnarray}
where $q$ is the relative momentum, $\omega$ is the collision energy
in the $N\Delta$ center of mass frame, and $P$ is the total momentum.
In the nonrelativistic baryon 
approximation,
we have
$$
P=\mid \vec{p}_{\Delta}+\vec{p} \mid \, ,
$$
$$
q = \mid \frac{m^0_\Delta\vec{p}-m_N\vec{p}_{\Delta}}{m^0_{\Delta}+m_N} \mid \, ,
$$
$$
\omega (\epsilon_{\Delta},\vec{p}_{\Delta},\vec{p}) = \epsilon_{\Delta}
+\epsilon_N(\vec{p}) - m^0_{\Delta} - m_N -\frac{(\vec{p}_{\Delta}+\vec{p})^2}
{2(m^0_{\Delta}+m_N)} \, ,
$$
where $m^0_\Delta=1280$ MeV is the bare mass of the $\Delta$ determined from 
fitting the $\pi N$ phase shifts in $P_{33}$ channel.

The equations for the G-matrix are identical to Eqs.\ (6)-(9)
except that the propagators are modified. In the partial-wave representation,
each equation has the following structure 
\begin{eqnarray}
G^{JT}_{\alpha,\delta}(q^\prime,&q&,\omega,P) = V^{JT}_{\alpha,\delta}
(q^\prime ,q,\omega ) \nonumber \\
&+&\sum_{\gamma} \int q^{\prime\prime 2}  
d q^{\prime\prime}\frac{V^{JT}_{\alpha,\gamma}(q^{\prime},q^{\prime\prime},
\omega )Q_{\gamma}(q^{\prime\prime},P,p_F)}{\omega - W_{\gamma}(q^{\prime\prime},P)
+ i \epsilon}
G^{JT}_{\gamma,\delta}(q^{\prime\prime},q,\omega,P)
\end{eqnarray}
where the greek subindices denote collectively 
the particle channels, $NN$ or $N\Delta$,
and the orbital and spin quantum numbers. In the propagator of Eq.\ (57),
$Q_{\gamma}(q'',P,p_F)$ is the angle-average Pauli operator \cite{lee4} and
\begin{eqnarray}
W_{\gamma}(q,P) = \epsilon_N(\bar{p}) + \epsilon_N(\bar{p})
\end{eqnarray}
for $\gamma = NN$ channel, and 
\begin{eqnarray}
W_{\gamma}(q) = \epsilon_N(\bar{p}_N) + \epsilon_{\Delta}(\bar{p}_{\Delta})
\end{eqnarray}
for $\gamma = N\Delta $ channel.
The single particle energies in the above equations are calculated by using
the angle-average momenta
$$\bar{p} = \sqrt{q^2 + \frac{1}{4}P^2},$$
and
$$ \bar{p}_N = \sqrt{(\frac{m_N P}{m^0_{\Delta}+m_N})^2 +q^2},  $$
$$ \bar{p}_{\Delta} = \sqrt{(\frac{m_{\Delta}P}{m^0_{\Delta} + m_{N}})^2
 +q^2}. $$

The nucleon single particle energy $\epsilon_N(p)$  
is taken from a variational many-body calculation by Wiringa \cite{wiringa}.
The single particle energy for the $\Delta$
is determined by the following self-consistency condition
\begin{eqnarray}
Re(\epsilon_\Delta (\vec{p}_\Delta)) = 
m^0_\Delta + \frac{p^2_\Delta}{2 m^0_{\Delta}} + 
Re(\Sigma^{(1)}(\vec{p}_{\Delta}, \epsilon_{\Delta}(\vec{p}
_{\Delta})) + Re(U_{\Delta}(\vec{p}_\Delta))) \, ,
\end{eqnarray}
where we have defined the $\Delta$ potential
\begin{eqnarray}
U_{\Delta}(\vec{p}_\Delta) =
\Sigma^{(2)}(\vec{p}_{\Delta},\epsilon_{\Delta}(\vec{p}_{\Delta})) \, .
\end{eqnarray} 

The medium effects on the cross sections of binary collisions defined 
in section III
can be included by replacing the scattering t-matrix in Eqs.\ (45), (48), (50), 
and (51) by the G-matrix. The collision energy is calculated from
real parts of the single particle energies $\epsilon_N(p)$ of 
Ref.\ \cite{wiringa} and $\epsilon_\Delta(p_\Delta)$ defined by the
self consistent condition Eq.\ (60).

\section{Results and Discussions}

We have applied the formula developed in section III to investigate the cross
sections of binary collisions involving a $\Delta^\ast$ excitation 
in either the entrance channel or
the exit channel. In Fig.\ 3, we show that the $NN \rightarrow N\Delta^\ast$
cross sections (dashed curve) calculated
by using the factorized form Eqs.\ (44)-(45) can reproduce reasonably well
the exact calculation of $NN \rightarrow \pi NN$ using Eqs.\ (35) and (40).
This suggests that the factorized forms defined by Eqs.\ (44)-(52) are sufficient for
investigating the main features of relativistic heavy-ion collisions.
The calculated total cross sections for all processes involving $\Delta^\ast$ 
are compared in Fig.\ 4. They have quite different energy dependencies.
The difference between $NN \rightarrow N\Delta^\ast$ and $N\Delta^\ast \rightarrow
NN$ is due to the flux factors $1/k_0^2$ of Eq.\ (45) and $1/\bar{k}^2$of 
Eq.\ (48). This can be understood from Fig.\ 5 in which 
the calculated $\bar{k}$ for the $N\Delta^\ast$ channel is compared with
the momentum
$k_0$ of the $NN$ channel. The calculated mass of the $\Delta^\ast$ is also 
displayed there. We see that as the collision energy increases the produced
$\Delta^\ast$ becomes heavier and moves faster.

The predicted differential cross sections at several energies
are compared in Fig.\ 6.
We see that the angular distributions
for $NN \rightarrow N\Delta^\ast$ is quite different from that of
$N\Delta^\ast \rightarrow NN$.
This is due to the fact that the important partial waves 
in $NN$ and $N\Delta$ are
quite different. For example, in the J=2 channel the outgoing
particles are in $^1D_2$ for the $N\Delta^\ast \rightarrow NN$, but
are in $^5S_2$ for the $NN\rightarrow N\Delta$.
Clearly such a dynamical difference in differential cross sections 
can not be easily obtained by
using detailed balance \cite{bertsch1,bertsch2,mosel2} 
to relate $N\Delta^\ast \rightarrow NN$ to the data of the $NN\rightarrow 
NN\pi $ reaction. We further notice that the
predicted 
$N\Delta^\ast \rightarrow N\Delta^\ast$ differential cross sections are forward
peaked.  This is also quite different from the usual assumption that
$N\Delta^\ast \rightarrow N\Delta^\ast$ is 
identical to $NN \rightarrow NN$ apart from some
isospin factors. The results shown in Figs.\ 4 and 6 are the consequences of
the considered meson-exchange mechanisms.

To get some ideas about the medium effects on collision cross sections, we 
assume that the mean field effect on the $\Delta$ propagation is 
the potential $U_\Delta$ defined in Eqs.\ (56) and (61). 
The calculated $U_\Delta$ at several densities are displayed in Fig.\ 7. Both the 
real parts and imaginary parts depend strongly on the density and the momentum.
Qualitatively, our results indicate that the $\Delta$ moves almost
freely in nuclear medium at high momentum, but is slowed down considerably
and is easily annihilated at low momentum. It will be interesting to explore the
consequence of the predicted $U_\Delta$ in determining the pion yields
in relativistic heavy-ion collisions.

The predicted $U_{\Delta}(p)$ is used to evaluate 
the $\Delta$ single particle energy Eq.\ (60) and the G matrix elements
which are the inputs to the calculations of the cross sections in medium, 
as explained in section IV. 
We have found that the medium effects are most dramatic in the region
where the collision energies are close to the $\pi NN$
threshold in free space.
This can be illustrated in a calculation where a $\Delta^\ast$ with a mass of
1236 MeV propagates under the influence of the mean field $U_\Delta$ 
and collides with a nucleon at rest in nuclear matter. 
The predicted dependence of the $N\Delta^\ast \rightarrow NN$
and $N\Delta^\ast \rightarrow N\Delta^\ast$ total cross sections 
on the density and the momentum is displayed in Fig.\ 8. 
There are two medium effects. The first one is the
mean field effect on the propagators of the G-matrix in Eq.\ (57).
This effect is found to be not very large in the momentum region considered.
The main medium effect is due to 
the change of the collision energies by the $\Delta$ potential $U_\Delta$ 
in the entrance channels. 
Since the potential energy(real part) becomes more attractive as density
increases (Fig.\ 7), the
collision energy for a given momentum of the incident $\Delta$ is shifted
to the lower energy region closer to the pion production threshold 
where the cross sections vary rapidly as shown in Fig.\ 4. 
We therefore see in Fig.\ 8 that the $N\Delta^\ast$ collision
cross sections in the low momentum region depend strongly on the density, 
and are greatly suppressed at high densities.  A similar situation
is also found for $NN\rightarrow N\Delta^\ast$, as illustrated in the lower half of
Fig.\ 9. The corresponding medium effects on the $NN\rightarrow NN$ are less
dramatic since the $NN$ elastic cross sections do not vary so rapidly in
the energy region near pion production threshold.

\section{Summary}

The binary collisions involving a $\Delta$ resonance in either the entrance
channel or the exit channel have been investigated within a Hamiltonian
formulation of $\pi NN$ interactions \cite{lee1,lee2,lee3}.  An averaging
procedure has been developed to define the experimentally measured
$NN\rightarrow \pi NN$ cross section in terms of an effective $NN\rightarrow
N\Delta^\ast$ cross section. In contrast to previous works
\cite{bertsch2,mosel2}, the main feature of the present approach is that the
mass and the momentum of the produced $\Delta^\ast$ at each collision energy are
calculated dynamically from the bare $\Delta \leftrightarrow \pi N$ vertex
interaction of the model Hamiltonian and are constrained by the unitarity
condition.  The procedure is then extended to define the effective cross
sections for the experimentally inaccessible $N\Delta^\ast \rightarrow NN$ and
$N\Delta^\ast \rightarrow N\Delta^\ast$ reactions. The calculations have been
performed to predict the energy dependencies and angular distributions of
these cross sections.  

The $\Delta$ mean field in nuclear matter has been
calculated by using the Bruckner-Hartree-Fock approximation developed in
\cite{lee4}. It is found that the $\Delta$ moves almost freely at high
momenta, but is slowed down considerably and get easily absorbed at low momenta. By
including the medium effect on the $\Delta$ propagation, the dependence of
the effective cross sections of the $N\Delta^\ast$ collisions on the density has
been investigated.  It is demonstrated that the density dependence is most
dramatic in the energy region close to the pion production threshold.

Our results can be used in the relativistic heavy-ion 
calculations using transport equations
\cite{bertsch1,bertsch2,mosel2,mosel3,li,ko,muntz}. To make further progress
in investigating $\Delta$-rich matter, it is necessary to investigate
the $\Delta^\ast \Delta^\ast$ collisions and the effects due to the higher mass
$N^\ast$ resonances. This requires the extension of the $\pi NN$ model Hamiltonian
defined by Eqs.\ (4)-(5) to include the two$-\pi$ channels and the 
development of 
a scattering theory with $\pi\pi NN$ unitary condition. Such a Hamiltonian model
can be accurately constructed only when the dynamics of our present understanding of
the $N^\ast$ excitations in 
$\pi N$ and $\gamma N$ reactions is  improved quantitatively.
The proposed pion facility at GSI \cite{gsi} and the $N^\ast$ program 
at CEBAF are essential for making progress in this direction. 

This work is supported by the U.S. Department of Energy, Nuclear Physics Division,
under contract W-31-109-ENG-38.


\newpage
\begin{figure}
\caption{The polarization observables $C_{LL}$ and $C_{LS}$ of np scattering
predicted by the $\pi NN$ model \protect\cite{lee1,lee2,lee3} are compared
with the recent data \protect\cite{spinka}.}
\end{figure}

\begin{figure}
\caption{Graphical representation of the $NN \rightarrow N\Delta \rightarrow
\pi NN$ reaction.}
\end{figure}

\begin{figure}
\caption{The calculated $NN \rightarrow \pi NN$ (solid curve) and 
$NN\rightarrow N\Delta^\ast$ (dotted curve) total cross sections 
are compared. The data are from the 
compilation of Ref.\ \protect\cite{arndt}.}
\end{figure}

\begin{figure}
\caption{The predicted total cross sections for the binary collisions involving
a $\Delta^\ast$ in either the entrance channel or the exit channel are compared.}
\end{figure}

\begin{figure}
\caption{The mass $\bar{M}^\ast$ and momentum $\bar{k}$ of 
the $\Delta^\ast$ in the $NN \rightarrow N\Delta^\ast$ reaction. $E_L$ is the
incident nucleon energy in the laboratory frame.}
\end{figure}

\begin{figure}
\caption{The predicted angular distributions evaluated at energies (in terms
of the corresponding nucleon laboratory energies) $E_L = $400 MeV (dotted
curves), 700 MeV (solid curves), and 900 MeV (dashed curves).}
\end{figure}

\begin{figure}
\caption{The calculated $\Delta$ potentia $U_{\Delta}(p_\Delta)$ at several
densities. The density $\rho$ is in units of $\rho_0 = 0.16$ nucleons-fm$^{-3}$.}
\end{figure}

\begin{figure}
\caption{The momentum-dependence of the calculated total cross sections
of $N\Delta^\ast \rightarrow NN$ and $N\Delta^\ast \rightarrow N\Delta^\ast$ collisions 
in nuclear matter. The density $\rho$ is in units of $\rho_0=0.16$ 
nucleons-fm$^{-3}$.}
\end{figure}

\begin{figure}
\caption{The momentum-dependence of the calculated $NN\rightarrow NN$ and 
$NN\rightarrow N\Delta^\ast$ collisions in nuclear matter. The density $\rho$ is in
unit of $\rho_0 = 0.16$ nucleons-fm$^{-3}$.}
\end{figure}

\end{document}